\documentclass[aps,preprint,floatfix,showpacs,letterpaper]{revtex4-1}
\usepackage{color}
\usepackage{latexsym,amssymb,amsmath,gensymb}
\usepackage{graphicx,dcolumn,bm}
\usepackage{kec,bigints}
\usepackage{mathtools}
\begin{document}

\title{Some nonrenormalizable theories are finite}

\author{Kevin Cahill}
\email{cahill@unm.edu}
\affiliation{Department of Physics \& Astronomy,
University of New Mexico, Albuquerque, New Mexico 
87131, USA}
\affiliation{Physics Department, Fudan University,
Shanghai 200433, China} 
\affiliation{School of Computational Sciences,\\
Korea Institute for Advanced Study, Seoul 130-722, Korea}

\date{\today}

\begin{abstract}
Some nonrenormalizable theories are less singular than all renormalizable theories, and one can use lattice simulations to extract physical information from them.  This paper discusses four nonrenormalizable theories that have finite euclidian and minkowskian Green's functions. 
Two of them have finite euclidian action densities and describe scalar bosons of finite mass.   The space of nonsingular
nonrenormalizable theories is vast.
\end{abstract}

\maketitle

\section{Introduction
\label{Introduction}}

Euclidian Green's functions 
are ratios of path integrals 
\begin{equation}
G_e(x_1,\dots,x_n) \equiv 
\langle 0 | \mathcal{T} 
\left[\phi_e(x_1) \dots \phi_e(x_n)\right] | 0 \rangle
= \frac{\bigintsss \! 
\phi(x_1) \dots \phi(x_n) \,
e^{{}- S[\phi]} D\phi }
{\bigintsss \!  e^{{}- S[\phi]} D\phi }
\label {euclidian time-ordered product}
\end{equation}
weighted by a negative
exponential \( e^{{}- S[\phi]} \)
of the euclidian action 
\begin{equation}
S[\phi] = \int \! L(\phi) \, d^4x  .
\label {euclidian action}
\end{equation} 
They are mean values of fields 
\begin{equation}
\langle 0 | \mathcal{T} 
\left[\phi_e(x_1) \dots \phi_e(x_n)\right] | 0 \rangle
= \int \! \phi(x_1) \dots \phi(x_n) \,
P[\phi] \, D \phi
\label {mean values}
\end{equation}
in a  probability distribution
\begin{equation}
P[\phi] = e^{{}- S[\phi]} / N
\label {probability distribution}
\end{equation}
normalized by
\begin{equation}
N = \int e^{{}- S_e[\phi]}  D\phi  .
\label {N}
\end{equation}
The weight that the probability
distribution \( P[\phi] \) gives to large values
of the field determines how singular
the Green's functions are.
They become less singular
as the probability
of large field values decreases. 
\par
Many nonrenormalizable theories
are less singular than all 
renormalizable theories.
In fact, theories
must be singular in order to be renormalizable.
A theory of a scalar field in 4-dimensions, for instance,
is renormalizable only if the highest power
of the field is \( \phi^4 \),
and so the probability \( P[\phi] \) for the field
to assume a large value \( |\phi(x)| \) 
within a hypercube of edge \( a \) 
is something like \( \exp( {} -  a^4 \, |\phi(x)|^4 ) \)\@.
 This exponential is small
only if \( |\phi(x)| > 1/a \), and so
the Green's function
\( \la 0 | \phi(0) \phi(a \hat x) | 0 \ra \)
diverges as \( 1/a^2 \) as \( a \to 0 \)\@.
In a theory with \( \phi^n(x) \) in its action density,
the 2-point function diverges as
\( 1/a^{8/n} \) as \( a \to 0 \),
becoming less singular 
as \( n \) exceeds 4
where the theory becomes 
nonrenormalizable~\cite{Cahill2013}\@.
\par 
One can't apply ordinary perturbation theory
to these nonrenormalizable theories,
but one can use lattice methods, 
expansions in powers of \( \hbar \),
and functional integration to
extract physical information from them.
\par
\par
In a theory with a euclidian action density \( L(x) \)
that is infinite when the modulus of the field
exceeds \( M \),
the probability \( P[\phi] \)
of fields with \( |\phi(x)| > M \) vanishes,
and the Green's functions are finite,
a possibility first suggested by
Boettcher and Bender~\cite{Bender1990}\@.
A  theory of a scalar boson field
with euclidian action density
\begin{equation}
L_1 = \half \, (\p_\mu \phi(x))^2 + 
\half \, m^2 M^2 \, \lt( \frac{1}{1 - \phi^2(x)/M^2}
-1 \rt) \equiv \half \, (\p_\mu \phi(x))^2 + 
\half \, m^2 M^2 \, 
\sum_{n=1}^\infty \frac{\phi^{2n}(x)}{M^{2n}}
\label {L_1}
\end{equation}
has finite Green's functions in euclidian and Minkowski
space~\cite{Cahill2013}\@.
This theory is not
renormalizable, but it is
less singular than those that are. 
\par
I use lattice methods in 
section~\ref{Theories with finite Green's functions}
to discuss this theory
(\ref{L_1}) and a similar one
\begin{equation}
L_s = \half \, (\p_\mu \phi(x))^2 + 
m^2 M^2 \, \lt( \frac{1}{\sqrt{1 - \phi^2(x)/M^2}}
-1 \rt)
\equiv \half \, (\p_\mu \phi(x))^2 + V_s(x)  
\label {L_s 0}
\end{equation}
both of which have finite Green's functions.
We will see that in these theories
the mean value in the vacuum of 
the (dimensionless) euclidian action density
diverges quadratically
as \( 4/3(aM)^2 \) as the 
dimensionless lattice spacing \( aM \to 0 \),
while that of the free theory diverges quarticly
as \( 1/2(aM)^4 \)\@. 
By doing the relevant nongaussian functional integrals,
I show that at any point \( x \) the 
\(2 n\)-point function \( \la 0 | \phi^{2n}(x) | 0 \ra \)
in these theories is given by
the simple formula~\cite{Cahill2013}
\begin{equation}
 \la 0 | \phi^{2n}(x) | 0 \ra = \frac{M^{2n}}{2n+1} 
\label {prediction}
\end{equation}
and that the mean value of the potential energy
\( V_s(x) \) in the ground state of the theory \( L_s \) is 
\begin{equation}
\la 0 | V_s(x) | 0 \ra = \half \, m^2 M
\int_{-M}^M  \lt(\frac{1}{ \sqrt{1 - \phi^2(x) /M^2 }}
-1 \rt) \, d\phi(x) \\
= \half ( \pi - 2 ) \, m^2 M^2 .
\label {half(pi-2)}
\end{equation}
\par
In section~\ref{Theories with finite ground-state energies},
I use lattice methods to show that 
in the theory with euclidian action density
\begin{equation}
L_f = M^4 \, \lt(
\frac{1}{\sqrt{1 - M^{-4} \lt[ (\p_\mu \phi(x))^2 
+ m^2 \phi^2(x) \rt]}} - 1 \rt)
\label {L_f}
\end{equation}
and in the closely related theory
\begin{equation}
L_{f ,s} = M^4 \lt( \frac{1}{\sqrt{1 - \p_\mu \phi^2/M^4}} -1 \rt)
+  m^2 M^2 \lt( \frac{1}{\sqrt{1 - \phi^2/M^2}} - 1 \rt)  ,
\label {L_{f, s} 0}
\end{equation}
the mean value of the euclidian action density 
in the vacuum is
is finite and equal to \( 0.7120 \, M^4 \)
for the case \( m = M \)\@.
In section~\ref{Masses},
I use lattice methods to estimate
the physical masses of the bosons
of the four theories of 
sections~\ref {Theories with finite Green's functions}
and \ref{Theories with finite ground-state energies}\@.
The theories \( L_1 \) and \( L_s \) are unacceptable
because the physical masses
of their scalar bosons are infinite.
But the physical masses
of the scalar bosons of the theories
\( L_f \) and \( L_{f, s} \) are finite and are
approximately 
\( m_f \approx M \) and \( m_{f, s} \approx M/20 \) 
for the case \( m = M \)\@.
In Section~\ref{Speculations about confinement, gravity,
and fermions},
I propose some nonsingular nonrenormalizable
theories of higher-spin fields.
The suggested \(SU(3)\) gauge theories
are much closer to Wilson's compact 
version of that theory and may justify 
his compactification of the gauge fields.  
I also propose two theories of gravity
that are less singular than ordinary 
quantum gravity and two ways to handle fermions.

\section{Theories with finite Green's functions
\label{Theories with finite Green's functions}}

The existence of quantum field theories
with finite Green's functions 
was first suggested by Boettcher and 
Bender~\cite{Bender1990}\@.
On a lattice of spacing \( a \),
the euclidian action 
of the theory (\ref{L_1}) with
finite Green's functions is
a sum over all \( N^4 \) vertices \( v \) 
of the vertex action
\begin{equation}
   \begin{split}
S_1(v) = {}& a^4 \lt[
\frac{1}{4}   
\sum_{j=1}^4 
\left(\frac{\phi(v) - \phi(v \pm \hat j)}{a} \right)^2 
+ \half m^2 M^2 \lt( 
\frac{1}{1 - \phi^2(v)/M^2 }
- 1 \rt)  \rt] \\
 = {}& a^4 M^4 \lt[
\frac{1}{4}   
\sum_{j=1}^4 
\left(\frac{\varphi(v) - \varphi(v \pm \hat j)}{aM} \right)^2 
+ \half \frac{m^2}{ M^2 }\lt( 
\frac{1}{1 - \varphi^2(v) }
- 1 \rt)  \rt]  
\label {S of first theory}
    \end{split}
\end{equation}
in which the field
\( \varphi = \phi/M \) and the product \( aM \)
are dimensionless.
The \(\pm\) signs mean that we average
the forward and backward derivatives.
We get the functional integrals
of the continuum theory
by sending the lattice spacing \( a \to 0 \)
and the size of the lattice \( N \to \infty \)\@.
In the limit \( M \to \infty \),
the action \( L_1 \) of the theory
with finite Green's functions (\ref{L_1})
and its lattice action (\ref{S of first theory})
respectively reduce to those of the free theory
   \begin{equation}
L_{0} =  \half \, (\p_\mu \phi(x))^2 + 
\half \, m^2 \, \phi^2(x) 
\label {free theory}
\end{equation} 
and 
\begin{equation}
   \begin{split}
S_0(v) = {}& a^4 \lt[
\frac{1}{4}   
\sum_{j=1}^4 
\left(\frac{\phi(v) - \phi(v \pm \hat j)}{a} \right)^2
+ \half \, m^2 \, \phi^2(v) \rt]  \\
= {}& \half \, a^4 m^4 \lt[
\frac{1}{2}  
\sum_{j=1}^4 
\left(\frac{\varphi(v) - \varphi(v \pm \hat j)}{a \, m} \right)^2
+ \, \varphi^2(v) \rt]  .
\label {S of free theory}
    \end{split}
\end{equation}

\begin{figure}[htbp] 
   \centering
   \includegraphics[width=5.5in]{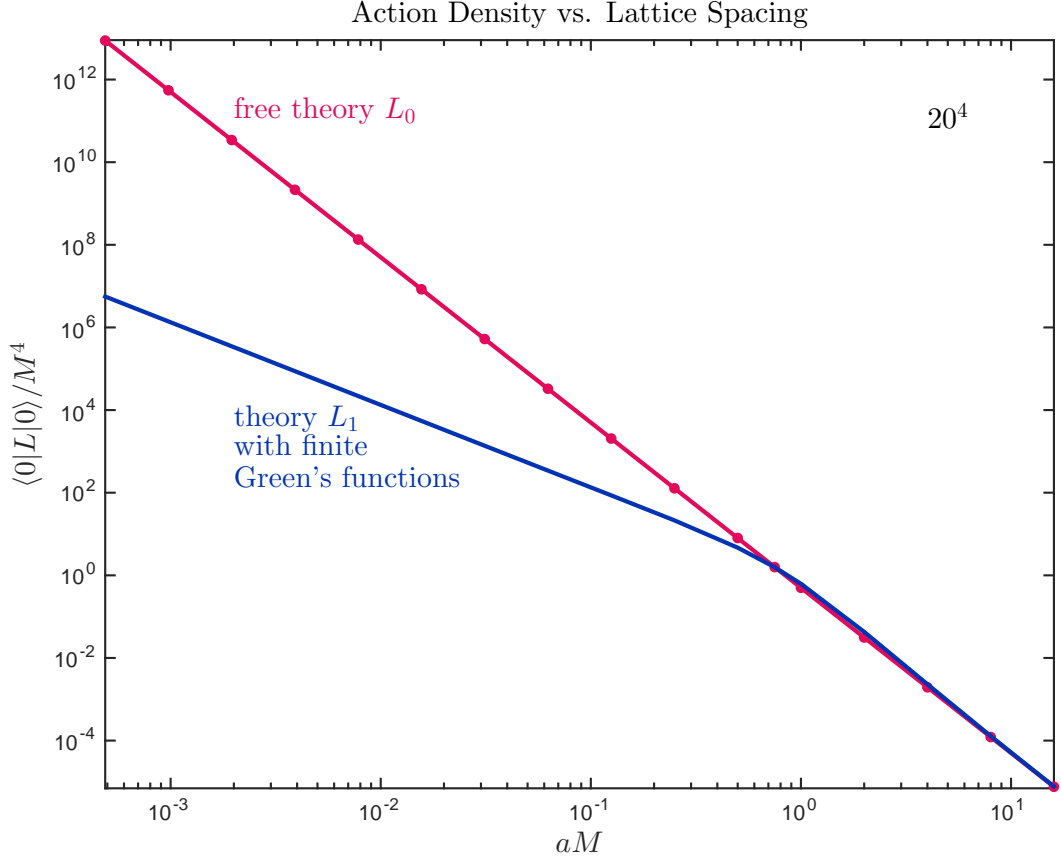} 
   \caption{The dimensionless
   ground-state euclidian
   action density \( \la 0 | L_1 | 0 \ra /M^4 \)
   of the theory \( L_1 \)
   with finite Green's functions
   (\ref{L_1}, solid blue line)
   and that of the free
   theory \( L_0 \) 
   (\ref {free theory}, solid dotted red line)
   are plotted against the 
   dimensionless lattice spacing \( aM \) from
   \( a M = 2^{-11} \) to \( a M = 2^4 \)
   for \( m = M \)\@.
   The two curves agree for \( a M > 2 \), but
   as  \( a M \to 0 \),
   the action density 
   of the theory \( L_1 \) 
   diverges quadratically as \( 4/(3 a^2) \)
   while that of the free theory \( L_0 \)
   diverges quarticly as \( 1/(2 a^4) \)\@.   }
   \label{fig:action density of first theory}
\end{figure}

\begin{figure}[htbp] 
   \centering
   \includegraphics[width=5.5in]{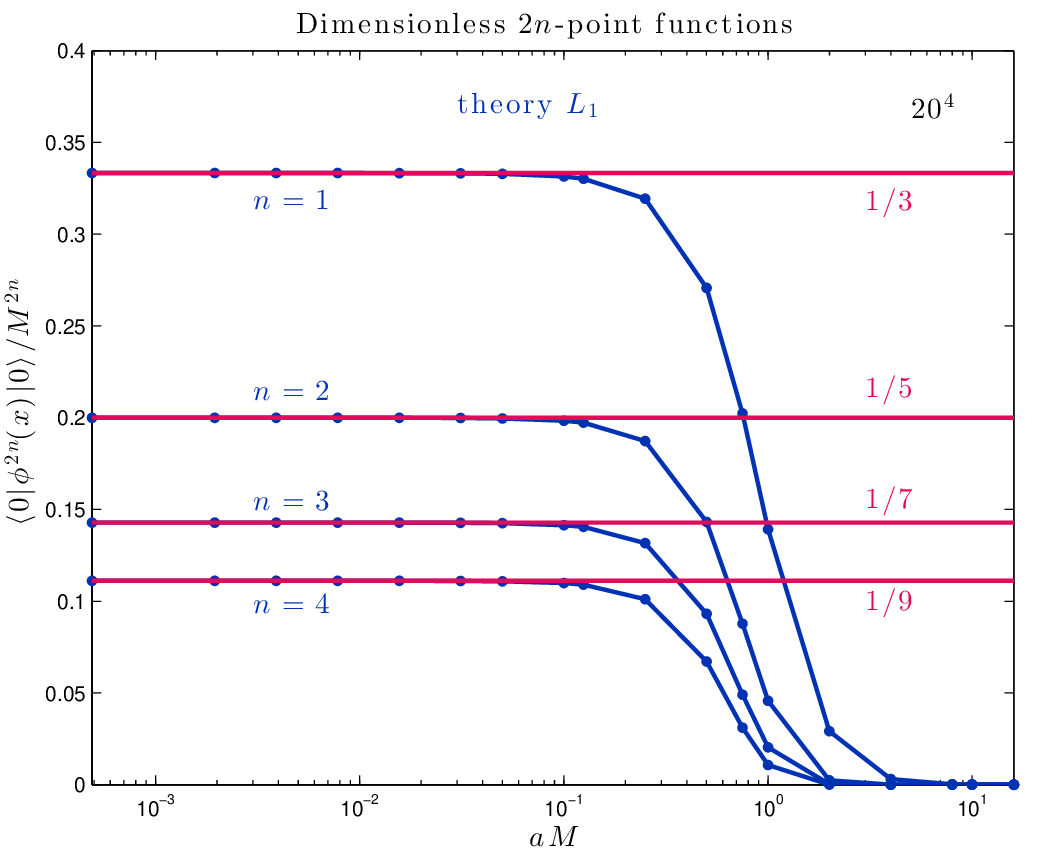} 
   \caption{In the theory (\ref{L_1})
   with finite Green's functions,
   the mean values 
   \( \la 0 | \phi^{2n}(x) | 0 \ra/M^{2n} \)
   (solid dotted blue lines)
   approach the fractions \( 1/(2n+1) \)
   (solid red lines)
   as the dimensionless lattice spacing \( a M \to 0 \)
   as predicted~\cite{Cahill2013}\@.}
    \label{fig:2n-point function}
\end{figure}  

\par
I have run Monte Carlo simulations~\cite{Creutz1983, *Cahill14.4}
with the action \( S_1 \) (\ref{S of first theory})
and \( S_0 \) (\ref{S of free theory}) on a \( 20^4 \) lattice
with periodic boundary conditions.
In all the simulations of this paper,
I allowed the fields to thermalize
for a million sweeps and then took data
in several runs of \( 2 \times 10^6 \) sweeps
for each set of parameters.
I restricted all the simulations 
to  the equal-mass case, \( m = M \),
because my computer resources were limited.
\par
Figure~\ref{fig:action density of first theory}
plots the dimensionless ground-state
   action density \( \la 0 | L_1 | 0 \ra /M^4 \)
   of the theory \( L_1 \)
   with finite Green's functions
   (\ref{L_1}, solid blue line)
   and that \( \la 0 | L_0 | 0 \ra /M^4 \) of the free
   theory 
(\ref {free theory}, solid dotted red line)
   against the 
   dimensionless lattice spacing \( aM \) from
   \( a M = 2^{-11} \) to \( a M = 2^4 \)
   for \( m = M \)\@.
   The two curves agree for \( a M > 2 \), but
   as  \( a M \to 0 \),
   the action density 
   of the theory \( L_1 \) 
   diverges quadratically as \( 4/(3 a^2) \)
   while that of the free theory \( L_0 \)
   diverges quarticly as \( 1/(2 a^4) \)\@.  
   The theory \( L_1 \)
   is less singular than the free theory \( L_0 \)\@.
      
   \begin{figure}[htbp] 
   \centering
   \includegraphics[width=5.5in]{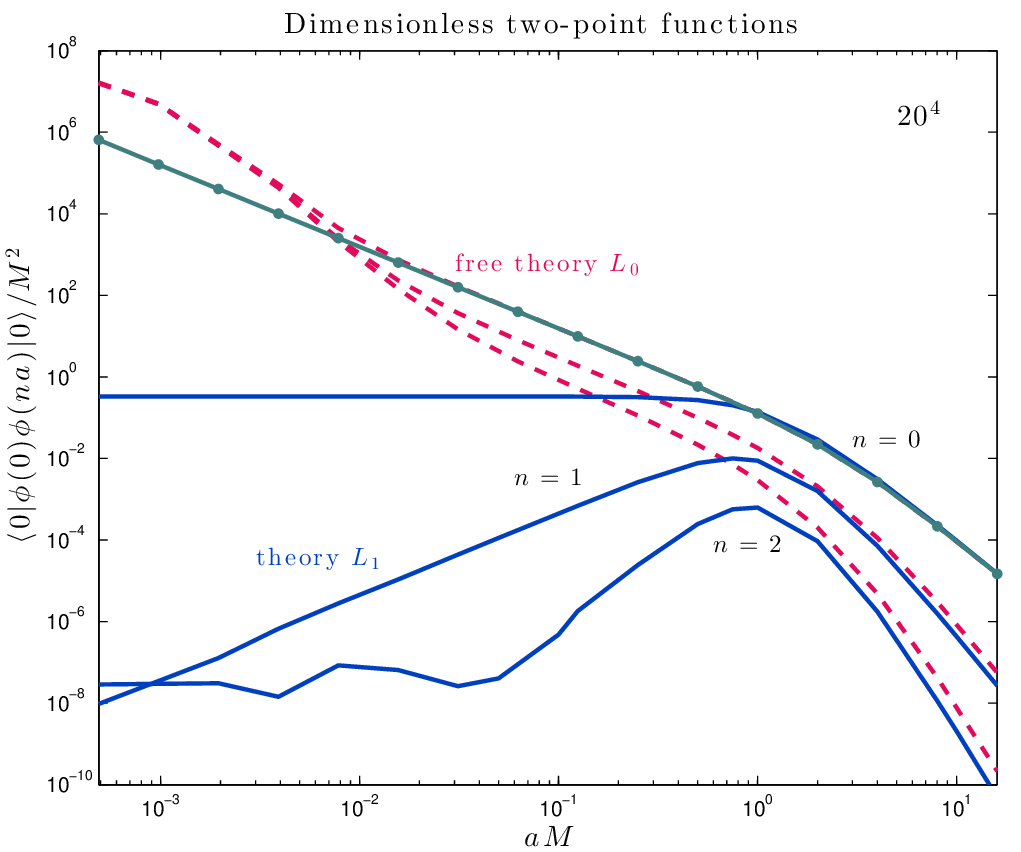} 
   \caption{The dimensionless 2-point function
   \( \la 0 | T \{ \phi(x) \phi(x + n a \hat j) \} |0\ra/M^2 \)
   (\ref {lattice 2-point function}) 
   of the theory \( L_1 \)
   with finite Green's functions (\ref{L_1}, solid blue lines)
   and of the free theory \( L_0 \)
   (\ref{free theory}, dashed red lines), 
   both on a \( 20^4 \) lattice, are plotted for \( n = 0 \),
   \( n = 1 \), and \( n = 2 \)  against
   the dimensionless
   lattice spacing \( am \) from
   \( am = 2^{-11} \) to \( a m = 2^4 \)
   for the case \( m = M \)\@.  
   Also plotted (for \( n = 0 \))
   is the exact 2-point function 
   of the free theory \( L_0 \) (\ref{free theory}) 
   on an infinite lattice (\ref{infinite lattice}, 
   solid dotted green line)\@.}
    \label{fig:Green Lite}
\end{figure}  

\par
The lattice action density
(\ref{S of first theory})
vanishes in the limit \( a \to 0 \),
unless  \( | \phi(x) | \ge M \),
in which case it's infinite.
Thus, the field is limited
to \( | \phi(x) | < M \),
and in the ratio (\ref{2n-point function})
of path integrals that gives the mean value
\( \la 0 | \phi^{2n}(x) | 0 \ra/M^{2n} \),
the integrations over the
field at \( x' \ne x \) all cancel.
This mean value
is therefore a ratio of one-dimensional 
integrals~\cite{Cahill2013}
\begin{equation}
\la 0 | \phi^{2n}(x) | 0 \ra =
\frac{\bigintsss \! 
\phi^{2n}(x) \,
e^{{}- S[\phi]} D\phi }
{\bigintsss \!  e^{{}- S[\phi]} D\phi }
 = \frac{\bigintsss_{-M}^M \! 
\phi^{2n}(x) \, d\phi(x) }
{\bigintsss_{-M}^M \!   d\phi(x) }
= \frac{M^{2n}}{2n+1} .
\label {2n-point function}
\end{equation}
The mean value of an odd power vanishes
by symmetry.
The lattice simulations 
for \( m  = M \) shown 
in Fig.~\ref{fig:2n-point function}
verify these formulas. 
The solid red horizontal lines
are the fractions \(1/(2n+1) \)
for \( n = 1\), 2, 3, and 4; 
the lattice estimates 
of the dimensionless ground-state 2n-point functions
\( \la 0 | \phi^{2n}(x) | 0 \ra/ M^{2n} \)
(solid dotted blue lines)
rapidly converge to these lines 
as the dimensionless lattice spacing
\( a M \) falls below 1 and approaches 0\@.
\par
In these simulations, 
the dimensionless 2-point function is the average
of \( N \) measurements of products of fields
\begin{equation}
   \begin{split}
      \frac{ 1 }{ M^2 } \, 
      \la 0 | T\{\phi(x) \phi(x + n a \hat j) \}| 0 \ra  
      = {}&  \frac{1}{N M^2} \sum_{k=1}^N \, 
       \phi_k(v) \, \phi_k(v + n  \hat j)  \\
         = {}&   \frac{ 1 }{N} \sum_{k=1}^N \, 
       \varphi_k(v) \, \varphi_k(v + n  \hat j) .
       \label {lattice 2-point function}
   \end{split}
\end{equation}

\begin{figure}[htbp]  
   \centering
   \includegraphics[width=5.5in]{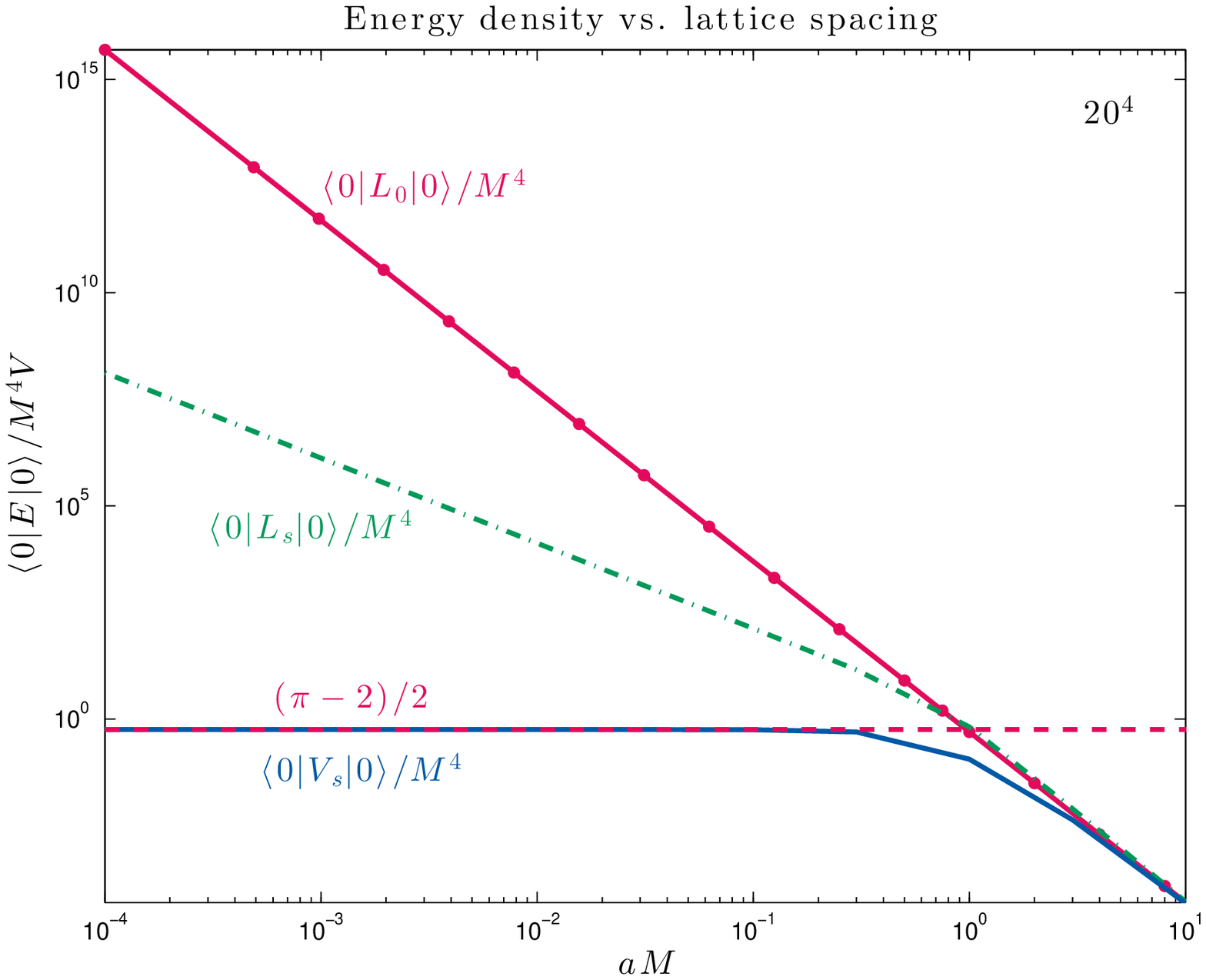} 
   \caption{The dimensionless, ground-state
   action density \( \la 0 | L_s | 0 \ra /M^4 \)
   (\ref{L_s}, dot-dash green line)
   and potential-energy density \(  \la 0 | V_s | 0 \ra /M^4 \)
   of the theory \( L_s \)
   (\ref{half(pi-2)}, solid blue line)
   and that \(  \la 0 | L_0  | 0 \ra /M^4 \)
   of the free theory  
   [(\ref{free theory}), solid dotted red line]
   are plotted against the 
   dimensionless lattice spacing \( aM \) from
   \( a M = 10^{-4} \) to \( a M = 10 \)
   for \( m = M \)\@.
   As \( a M \to 0 \), the potential-energy density 
   \(  \la 0 | V_s | 0 \ra /M^4 \)
   converges to the theoretical value
   \( (\pi -2)/2 \) [(\ref{half(pi-2)}), dashed red line].
   The action densities agree for \( a M > 2 \), but
   as  \( a M \to 0 \),
   the action density 
   of the theory \( L_s \) 
   diverges quadratically as \( 4/(3 a^2) \)
   while that of the free theory \( L_0 \)
   diverges quarticly as \( 1/(2 a^4) \)\@.}
   \label {fig:action density of Ls theory}
\end{figure}

\par
The exact dimensionless 2-point function 
of the free theory \( L_0 \) in the continuum is
\begin{equation}
   \begin{split}
 \frac{ 1 }{ M^2 }  \, 
 \la 0 | T\{\phi(x) \phi(x + n a \hat j) \}| 0 \ra  
      = {}& \frac{1}{4 \pi^2 n^2 a^2 M^2} \,
      n a M \, K_1(n a M) \\
      \approx {}& \frac{
      1 - \fourth\lt(1 - 2 \gamma + 2 \ln(2/naM) \rt) (naM)^2}
      {4 \pi^2 n^2 a^2 M^2}  \\
       \approx {}& \frac{1}{(2 \pi n a M)^{3/2}} \, e^{-naM} 
      \label {exact, free 2-point function} 
    \end{split}
\end{equation}   
in which the approximation 
of the second line holds  
for \( n a M << 1 \)
and that of the third for  \( n a M >> 1 \)\@.
The exact dimensionless 2-point function 
of the free theory \( L_0 \) on an infinite lattice is
\begin{equation}
   \begin{split}
\frac{ 1 }{ M^2 } \,
\la 0 | T\{\phi(x) \phi(x + n a \hat j) \}| 0 \ra 
      = {}& \frac{1}{a^2 M^2} \int_{-\pi/a}^{\pi/a}
      \frac{a^4 d^4q}{(2\pi)^4}
      \frac{\exp(i n a q_j)}
      {a^2 M^2 + 2 \sum_k (1 - \cos a q_k)} \\
= {}& \frac{1}{a^2 M^2} \int_{-\pi/a}^{\pi/a}
      \frac{a^4 d^4q}{(2\pi)^4}
      \frac{\exp(i n a q_j)}
      {a^2 M^2 + 4 \sum_k \sin^2 (aq_k/2)} \\
= {}& \frac{1}{a^2 M^2} \int_{-\pi}^{\pi}
      \frac{d^4p}{(2\pi)^4}
      \frac{\exp(i n p_j)}
      {a^2 M^2 + 4 \sum_k \sin^2 (p_k/2)} \\ 
= {}& \frac{1}{\pi^4 a^2 M^2 } \int_0^{\pi}
      \frac{\cos(n p_j)}
      {a^2 M^2 + 4 \sum_k \sin^2 (p_k/2)} \, d^4p  .                
   \end{split}
   \label {infinite lattice}
\end{equation}
 
\begin{figure}[htbp]  
   \centering
   \includegraphics[width=5.5in]{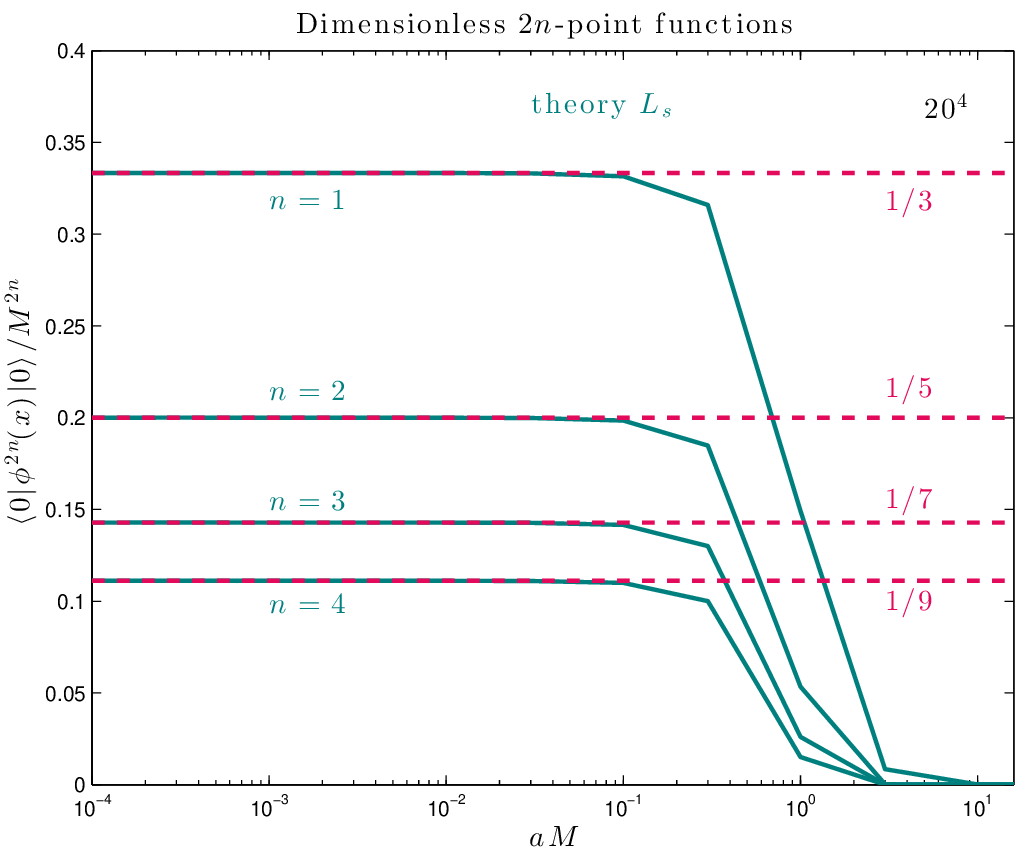} 
   \caption{In the theory (\ref{L_s}),
   which also has finite Green's functions,
   the mean values 
   \( \la 0 | \phi^{2n}(x) | 0 \ra/M^{2n} \)
   approach the fractions \( 1/(2n+1) \)
   as the dimensionless lattice spacing \( a M \to 0 \)\@.}    \label{fig:S2n-point function}
\end{figure}  

\par
For fields separated by  \( n =  0 \), 1, and 2
lattice spacings on a \( 20^4 \) lattice,
Fig.~\ref{fig:Green Lite} plots
the dimensionless lattice
2-point function 
 \( \la 0 | T \{ \phi(x) \phi(x + n a \hat j) \} |0\ra / M^2 \)
(\ref{lattice 2-point function}) for 
the theory with finite Green's functions
[(\ref{L_1}), solid blue lines]
and for the free theory 
[(\ref{free theory}), dashed red lines] 
against the dimensionless
   lattice spacing \( aM \) from
   \( aM = 2^{-7} \) to \( a M = 2^4 \)
   for the case \( m = M \)\@.  
   The two theories agree for 
   \( aM \ge 2 \)\@.
For separations of \( n = 0 \) lattice spacings,
Fig.~\ref{fig:Green Lite}
also plots 
the exact dimensionless 2-point function
   of the free theory \( L_0 \) (\ref{free theory}) 
   on an infinite lattice [(\ref{infinite lattice}), 
   solid dotted green line]
which reveals lattice artifacts for
 \( aM \le 2^{{}-5} \)\@.
 For \( aM < 0.03 \), the wobble in 
 \( \la 0 | T \{ \phi(x) \phi(x + 2 a \hat j) \} |0\ra / M^2 \)
 of the theory \( L_1 \) is due
 to insufficient statistics. 

 \par
\begin{figure}[htbp]  
   \centering  
   \includegraphics[width=5.5in]{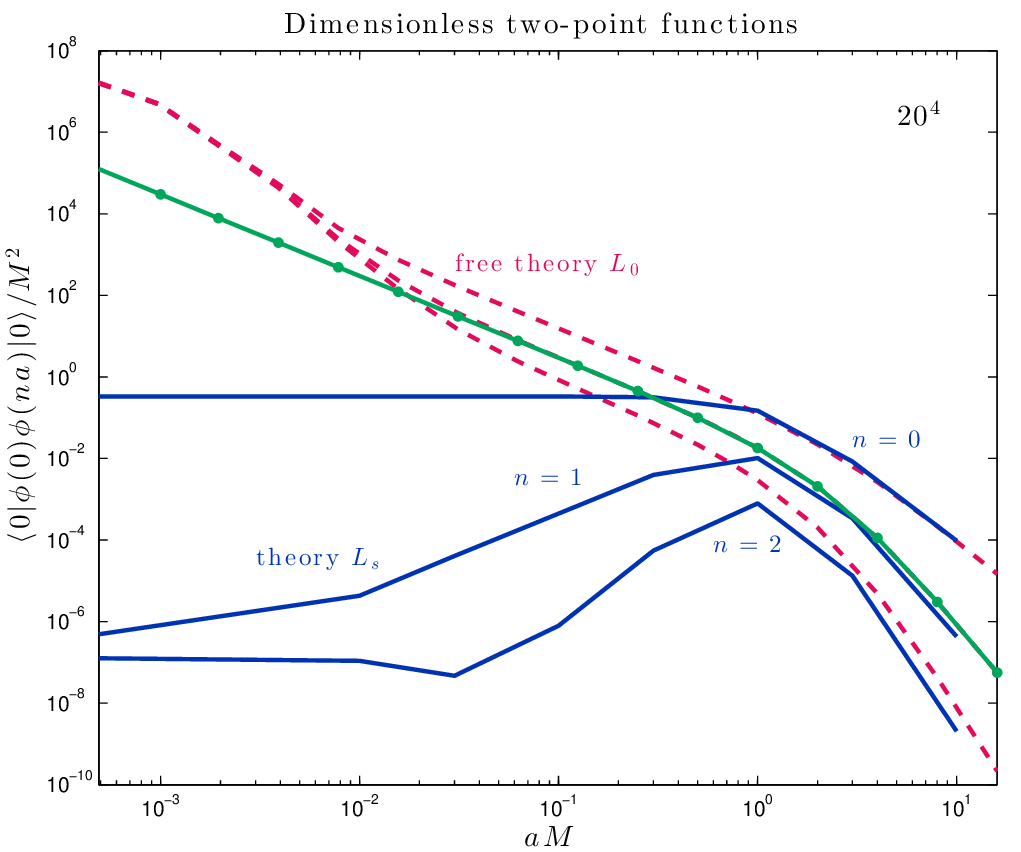} 
      \caption{The dimensionless 2-point function
   \( \la 0 | T \{ \phi(x) \phi(x + n a \hat j) \} |0\ra/M^2 \)
   (\ref {lattice 2-point function}) 
   of the theory \( L_s \)
   with finite Green's functions [(\ref{L_s}), solid blue lines]
   and of the free theory \( L_0 \)
   [(\ref{free theory}), dashed red lines], 
   both on a \( 20^4 \) lattice, are plotted for \( n = 0 \),
   \( n = 1 \), and \( n = 2 \)  against
   the dimensionless
   lattice spacing \( am \) from
   \( am = 2^{-11} \) to \( a m = 2^4 \)
   for the case \( m = M \)\@.  
   Also plotted (for \( n = 1 \))
   is the exact 2-point function
   of the free theory \( L_0 \) (\ref{free theory}) 
   on an infinite lattice 
   [(\ref{infinite lattice}), solid dotted green line]\@.}    
   \label {fig:S2-point function}
\end{figure}

We turn now to 
the theory (\ref{L_s 0}) with euclidian action density
\begin{equation}
L_s = \half \, (\p_\mu \phi(x))^2 + 
m^2 M^2 \, \lt( \frac{1}{\sqrt{1 - \phi^2(x)/M^2}}
-1 \rt)
\equiv \half \, (\p_\mu \phi(x))^2 + V_s(x)  .
\label {L_s}
\end{equation}
This theory also has finite Green's functions.
Arguments similar to
the ones that gave us the mean-value formulas
(\ref{2n-point function}) show
that the mean value 
the potential-energy density \( V_s \)
in the ground state is  
\begin{equation}
\la 0 | V_s(x) | 0 \ra = (2M)^{-1} m^2 M^2
\int_{-M}^M  \lt(\frac{1}{ \sqrt{1 - \phi^2(x) /M^2 }}
-1 \rt) \, d\phi(x) \\
= \half ( \pi - 2 ) \, m^2 M^2 ,
\label {half(pi-2)}
\end{equation}
that the mean value of
the \(2n\)th
power of the field at a given point \( x \) is
\begin{equation}
\la 0 | \phi^{2n}(x) | 0 \ra = \frac{M^{2n}}{2n+1} ,
\label {2n-point functions}
\end{equation}
while those of the odd powers vanish
by symmetry, and that the dimensionless action density
in the ground state \( \la 0 | L_s | 0 \ra / M^4 \)
diverges as \( 4/(3a^2) \)\@.
These predictions are verified
for \( m = M \) by the lattice simulations
displayed in Figs.~\ref{fig:action density of Ls theory},
\ref{fig:S2n-point function}, and
\ref{fig:S2-point function}\@.
The dimensionless 2-point function
   \( \la 0 | T \{ \phi(x) \phi(x + n a \hat j) \} |0\ra/M^2 \)
   (\ref {lattice 2-point function}) 
   of the theory \( L_s \)
   with finite Green's functions [(\ref{L_s}), solid blue lines]
   and of the free theory \( L_0 \)
   [(\ref{free theory}), dashed red lines], 
   both on a \( 20^4 \) lattice, are plotted 
   in Fig.~\ref{fig:S2-point function} for \( n = 0 \),
   \( n = 1 \), and \( n = 2 \)  against
   the dimensionless
   lattice spacing \( am \) from
   \( am = 2^{-11} \) to \( a m = 2^4 \)
   for the case \( m = M \)\@.  
   Also plotted (for \( n = 1 \))
   is the exact 2-point function
   of the free theory \( L_0 \) (\ref{free theory}) 
   on an infinite lattice 
   [(\ref{infinite lattice}), solid dotted green line]\@.

\section{Theories with finite ground-state action densities
\label{Theories with finite ground-state action densities}}

The euclidian action densities 
of the theories \( L_1 \) and \( L_s \)
(\ref{L_1} and \ref{L_s 0}) diverge
because their derivatives
contribute only quadratically
to their action densities.
We can make the mean value
of the ground-state euclidian action density finite by
using as the euclidian action density (\ref{L_f})
\begin{equation}
L_f = M^4 \, \lt(
\frac{1}{\sqrt{1 - M^{-4} \lt[ (\p_\mu \phi(x))^2 
+ m^2 \phi^2(x) \rt]}} - 1 \rt)
\label {L_f redux}
\end{equation}
or
\begin{equation}
L_{f ,s} = M^4 \lt( \frac{1}{\sqrt{1 - \p_\mu \phi^2/M^4}} -1 \rt)
+  m^2 M^2 \lt( \frac{1}{\sqrt{1 - \phi^2/M^2}} - 1 \rt)  .
\label {L_{f, s}}
\end{equation}
\par
The euclidian action on a lattice of spacing \( a \) 
of the theory (\ref{L_f redux}) is
a sum over all \( N^4 \) vertices \( v \) 
of the vertex action
\begin{equation}
   \begin{split}
S_f(v) = {}& a^4 M^4 \lt\{ 1 - 
M^{-4} \lt[
\frac{1}{2}   
\sum_{j=1}^4 
\left(\frac{\phi(v) - \phi(v \pm \hat j)}{a} \right)^2
+ m^2 \phi^2(v) \rt] \rt\}^{-1/2}  - a^4 M^4 \\
 = {}& a^4 M^4 \lt\{ 1 - 
\lt[
\frac{1}{2}   
\sum_{j=1}^4 
\left(\frac{\varphi(v) - \varphi(v \pm \hat j)}{aM} \right)^2
+ \frac{m^2}{M^2} \varphi^2(v) \rt] \rt\}^{-1/2} 
- a^4 M^4  
\label {S lattice of L_f}
    \end{split}
\end{equation}
in which the field
\( \varphi = \phi/M \) and the product \( aM \)
are dimensionless.
The \(\pm\) signs mean that we average
the forward and backward derivatives.

\begin{figure}[htbp]  
   \centering
   \includegraphics[width=5.5in]{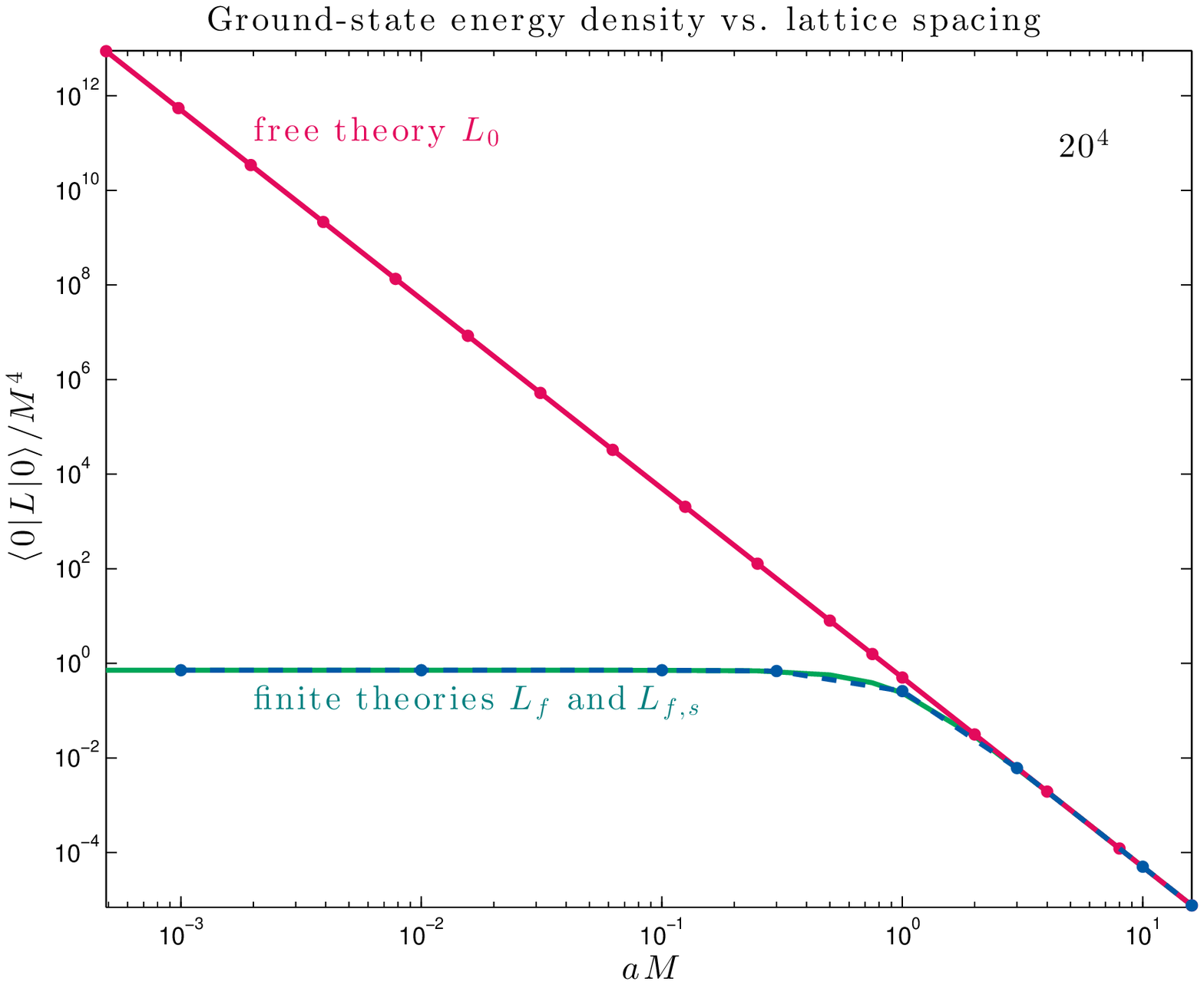} 
   \caption{The ground-state dimensionless
   action densities \( \la 0 | L_{f} | 0 \ra /M^4 \) 
   and \(  \la 0 | L_{f, s}  | 0 \ra /M^4 \)
   of the finite 
   theories [(\ref {L_f}), solid green line]
   and [(\ref{L_{f, s}}), dashed-and-dotted blue line]
   and of the free theory \( \la 0 | L_0  | 0 \ra  / M^4 \) 
   [(\ref {free theory})--(\ref{S of free theory}), 
   solid dotted red line]
   on a \(20^4\) lattice
   are plotted against the 
   dimensionless lattice spacing \( aM \) 
   for \( m = M \)\@.   
   As  \( a M \to 0 \),
   the action densities 
   of the finite theories approach 
   \( 0.7120 \, M^4 \), while that 
   of the free theory \( L_0 \)
   diverges quarticly as \( 1/(2 a^4) \)\@.   }
   \label{fig:action density}
\end{figure}

\begin{figure}[htbp]  
   \centering
   \includegraphics[width=5.5in]{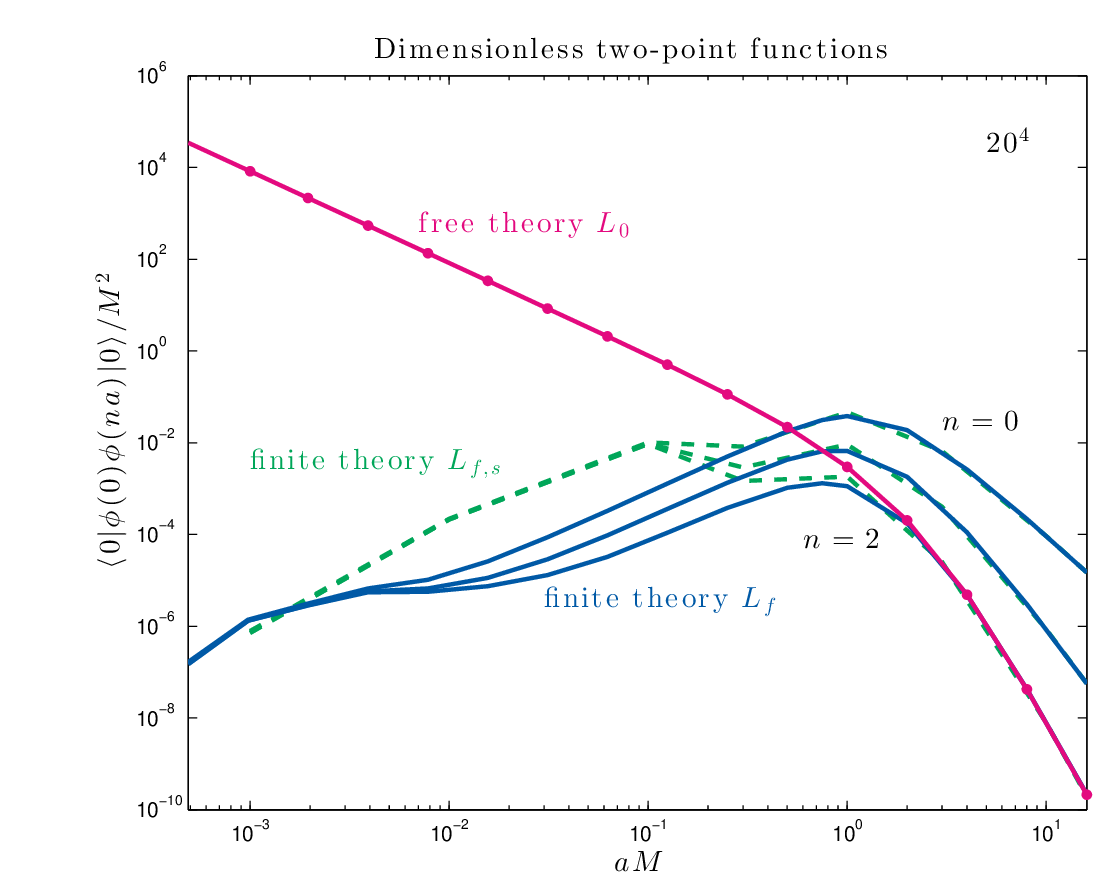} 
   \caption{The dimensionless 2-point functions
   \( \la 0 | T \{ \phi(x) \phi(x + n a \hat j) \} |0\ra/M^2 \)
   (\ref {lattice 2-point function}) of the finite 
   theories \( L_f \)  [(\ref{L_f}), solid blue lines]
   and \( L_{f, s} \) [(\ref{L_{f, s}}), dashed green lines]
   on a \( 20^4 \) lattice are plotted for \( n = 0 \),
   \( n = 1 \), and \( n = 2 \)  against
   the dimensionless
   lattice spacing \( aM \) from
   \( aM = 2^{-11} \) to \( a M = 2^4 \)
   for the case \( m = M \)\@.  
   Also plotted (for \( n = 2 \))
   is the exact 2-point function
   of the free theory \( L_0 \) (\ref{free theory}) 
   on an infinite lattice [(\ref{infinite lattice}), 
   solid dotted red line]\@.
   }
   \label {fig:free 2-point function}
\end{figure}

\par
We recover functional integrals like
\begin{equation}
\la 0 | L_{f} | 0 \ra  = \frac{\displaystyle{\int}
L_{f}
\, \exp \left[ - \int \!
L_{f}(\phi) \, d^4x \rt] D\phi }
{\displaystyle{\int} 
\exp \left[ - \int \!
L_{f}(\phi) \, d^4x \right] D\phi}  .
\label {dark}
\end{equation}
by taking the twin limits \( a \to 0 \)
and \( N \to \infty \)\@.

\par
I have run Monte Carlo simulations
with the lattice action (\ref {S lattice of L_f})  
of the finite theory \( L_f \)
and with the lattice action of the 
closely related theory \( L_{f, s} \) (\ref{L_{f, s}})
on a \( 20^4 \) lattice
with periodic boundary conditions
for the equal-mass case, \( m = M \)\@.
Figure~\ref{fig:action density} 
plots the mean values (\ref{dark}) in the ground state  
of the dimensionless action densities
\( \la 0 | L_{f} | 0 \ra/M^4 \)
[(\ref{L_f}), solid green line]
and \( \la 0 | L_{f, s} | 0 \ra/M^4 \)
(dashed-and-dotted blue line)
as well as that of the free theory 
\( L_0 \) [(\ref {free theory}), solid dotted red line]
for values of the lattice spacing running from
\( a = 2^{-11}/M \) to \( a = 2^4 / M \)\@.
The three curves agree for \( a > 2/M \)\@. 
But as  \( a M \to 0 \),
   the action densities \( \la 0 | L_f | 0 \ra \)
   and  \( \la 0 | L_{f, s} | 0 \ra \)
   approach \( 0.7120 \, M^4 \), 
   while  \( \la 0 | L_0 | 0 \ra \) diverges quarticly
as \( 1/(2 a^4) \)\@. 
Incidentally, the  ground-state action density 
\( 0.7120 \, M^4 \) of \( L_f \) and \( L_{f, s} \)
would fit the experimental value 
\( (0.00224
\,\, \mbox{eV})^4 \) of the
dark-energy density~\cite{PlanckCosmological, *Cahillpop-ph13}
if we set \( m = M = 2.44 \) meV\@.

\par
For fields separated by  \( n =  0 \), 1, and 2
lattice spacings on a \( 20^4 \) lattice,
Fig.~\ref{fig:free 2-point function} plots
the dimensionless lattice
2-point function 
 \( \la 0 | T \{ \phi(x) \phi(x + n a \hat j) \} |0\ra / M^2 \)
(\ref{lattice 2-point function}) for 
the finite theory \( L_f \) [(\ref{L_f}), solid blue lines]
and for the similar theory \( L_{f, s} \)
[(\ref{L_{f, s}}), dashed green lines]
against the dimensionless
   lattice spacing \( aM \) from
   \( aM = 2^{-11} \) to \( a M = 2^4 \)
   for the case \( m = M \)\@.  
The two theories agree with each other
and with the free theory \( L_0 \) 
for  \( aM \ge 1 \), but 
as the dimensionless
lattice spacing \( aM \to 0 \),
the (bare) Green's functions of 
 \( L_f \)  and \( L_{f, s} \) approach zero.
Figure~\ref{fig:free 2-point function}
also plots for separations of \( n = 2 \) lattice spacings
the exact dimensionless 2-point function
   of the free theory \( L_0 \) (\ref{free theory}) 
on an infinite lattice 
[(\ref{infinite lattice}), solid dotted red line]\@.

\section{Masses
\label{Masses}}

\begin{figure}[htbp] 
   \centering
   \includegraphics[width=5.5in]{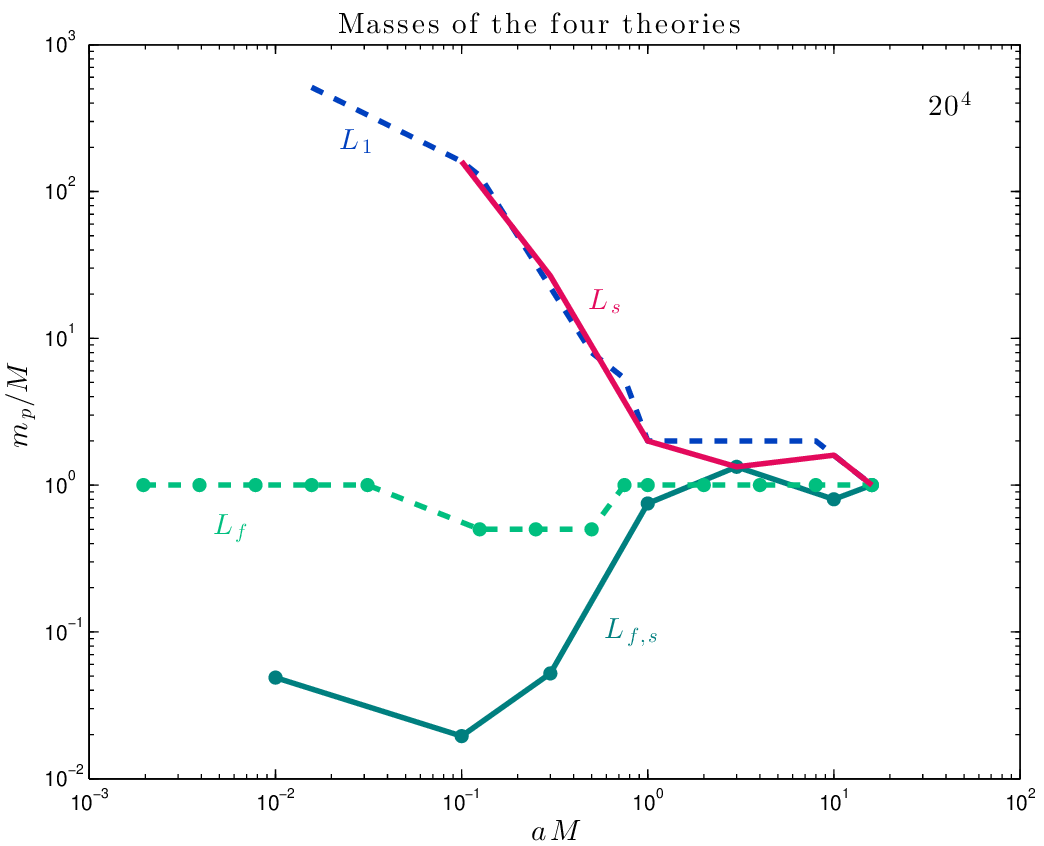} 
   \caption{At various values of
   the dimensionless lattice spacing \( aM \),
   the ratio of the physical mass \( m_p \)  
   (\ref{mass formula})  
   to the mass parameter \( M \)
   is plotted for the theories \( L_1 \) 
   [(\ref{L_1}), dashed blue line],
   \( L_s \) 
   [(\ref {L_s}), solid red line],
   \( L_f \)
   [(\ref{L_f}), dashed-and-dotted green line], 
   and \( L_{f, s} \)
   [(\ref{L_{f, s}}), solid dotted blue-green line]
   for the case  \( m = M \) \@.}
    \label{fig:masses}
\end{figure} 

One may use analytic and lattice methods
to estimate the physical masses
of the bosons of the four theories described in 
sections~\ref{Theories with finite Green's functions}
and ~\ref {Theories with finite ground-state energies}\@.
 \par
Although the theories \( L_1 \) and \( L_s \)
have finite Green's functions,
they describe bosons of infinite mass. 
We can see why the mass of the theory \( L_1 \) diverges
by considering the equation of motion
of its field \( \phi(x) \) in Minkowski spacetime
\begin{equation}
\ddot \phi(x) - \triangle \phi(x) 
= - m^2 \phi(x) \lt[ 1 - \phi^2(x) /M^2 \rt]^{-2} .
\label {equation of motion}
\end{equation}
I will approximate the nonlinear term
\( (1 - \phi^2/M^2)^{-2} \)
by its mean value
in the vacuum 
\begin{equation}
\ddot \phi(x) - \triangle \phi(x) 
= - m^2 \phi(x)  \,
\la 0 | \lt[ 1 - \phi^2(x) /M^2 \rt]^{-2} | 0 \ra .
\label {approximate equation of motion}
\end{equation}
An argument similar to
the one that gave us the mean-value formulas
(\ref{2n-point function}) shows that this mean value
diverges
\begin{equation}
   \begin{split}
      \la 0 | \lt[ 1 - \phi^2(x) /M^2 \rt]^{-2} | 0 \ra 
      = {}&  (2M)^{-1} \int_{-M}^M  \lt[ 1 - \phi^2 /M^2 \rt]^{-2} 
      \, d\phi \\
        = {}& (2M)^{-1} 
        \int_{-M}^M  \frac{1}{(1 - \phi /M)^2(1+\phi/M)^2}  
        \, d \phi = \infty .
   \end{split}
   \label{infinity}
\end{equation}
So the physical mass of the boson is infinite.
\par
Similarly, the equation of motion
of the field of the second theory \( L_s \)
in Minkowski space is
\begin{equation}
\ddot \phi(x) - \triangle \phi(x) 
= - m^2 \phi(x) \lt[ 1 - \phi^2(x) /M^2 \rt]^{-3/2} .
\label {equation of motion 2}
\end{equation}
I again use the approximation
\begin{equation}
\ddot \phi(x) - \triangle \phi(x) 
= - m^2 \phi(x)  \,
\la 0 | \lt[ 1 - \phi^2(x) /M^2 \rt]^{-3/2} | 0 \ra .
\label {approximate equation of motion 2}
\end{equation}
The mean value is infinite
\begin{equation}
   \begin{split}
      \la 0 | \lt[ 1 - \phi^2(x) /M^2 \rt]^{-3/2} | 0 \ra 
      = {}&  (2M)^{-1} \int_{-M}^M  \lt[ 1 - \phi^2 /M^2 \rt]^{-3/2} 
      \, d\phi = \infty ,
         \end{split}
   \label{infinity 2}
\end{equation}
and so is the mass of the boson.
Figure~\ref{fig:masses} verifies
these estimates of the masses
bosons of theories \( L_1 \) and \( L_s \)\@.
\par
The theories \( L_f \) and \( L_{f, s} \)
are much more complicated than 
\( L_1 \) and \( L_s \), and I don't
have a good analytic argument 
that shows that their masses are finite.
But these theories
bound the derivatives \( (\p_\mu \phi)^2 \)
as well as the fields \( \phi^2 \),
and the lattice action (\ref{S lattice of L_f})
shows that if the derivatives are to be bounded
in the limit \( a M \to 0 \),
then the mean values of the fields 
\( \la 0 | \phi^{2n}(x) | 0 \ra \) must become tiny
in that limit as a glance at Fig.~\ref {fig:free 2-point function}
reveals.  Thus, we expect the masses of the bosons
of the second pair of theories \( L_f \)
and \( L_{f, s} \) to be finite.
\par
I used the 2-point functions 
displayed in Figs.~\ref{fig:Green Lite}, \ref{fig:S2-point function},
and \ref{fig:free 2-point function}
to estimate the masses of the bosons
of the four theories as follows.
Let 
\begin{equation}
f(n a M) \equiv \frac{
\la 0 | T\{\phi(x) \, \phi(x + n a \hat j) \} | 0 \ra}
{M^2}
\label {f(n a M)}
\end{equation}
be the dimensionless 2-point function
for one of the four theories
simulated on a \( 20^4 \) lattice,
and let \( f_0(n a M) \) be the same thing
for the free theory \( L_0 \)\@.
For each theory, and each value
of the dimensionless lattice spacing
\( a M \),
I minimized the sum
\begin{equation}
\sum_{n=0}^1\lt( \frac{ f((n+1) a M) }{ f(n a M) } 
- \frac{ f_0( (n+1) (a M)') } { f_0(n (a M)') } \rt)^2
\label {what I minimized}
\end{equation}
over values of \( (aM)' \) 
from \( (aM)' = 2^{-11} \) to \( (aM)' = 2^4 \)
at which I had measured the
2-point function of the free theory
on a \( 20^4 \) lattice.
I took the upper limit on 
the sum over \( n \) to be unity
rather than 2 or more in order
to stay within a range in which
my statistical errors were small.
I then estimated the physical mass 
of the boson to be the limit
\begin{equation}
\lim_{aM \to 0} m(aM) \equiv
\lt( \lim_{aM \to 0} \frac { (aM)' }{aM} \rt) \, M .
\label {mass formula}
\end{equation}
The values of the mass ratios
from my simulations of the four theories
are displayed in Fig.~\ref{fig:masses}\@.
This figure shows that the masses
of the bosons of the first two theories
\( L_1 \) and \( L_s \) diverge as \( aM \to 0 \),
but that those of the bosons of the third and fourth
theories are finite with
\( m_f \approx M \) and
\( m_{f, s} \approx M/20 \)\@.

\section{Speculations about confinement, gravity,
and fermions
\label {Speculations about confinement, gravity,
and fermions}}

\par
Many nonrenormalizable theories
are less singular than that of a free field.
The space of such theories is vast.
We can make a typical theory
of scalar and vector bosons less
singular by replacing its 
euclidian action density \( L \) 
by~\cite{Cahill19.1}
\begin{equation}
L' = \frac{L}{1- L/M^4} \qquad
\mbox{or by} \qquad
L'' = M^4 \lt[ \exp(L/M^4) -1 \rt] 
\label {L'}
\end{equation}
or by any expression
that grows dramatically for large \( L \)\@. 

\subsection{Confinement
\label{Confinement}}

The euclidian action density 
of  \( SU(3) \) gauge theory 
(without fermions and \(\theta\)-vacua)
is the trace
\begin{equation}
L_3 = \frac{1} {2 g^2} \, \tr 
\lt( F^2_{\mu \nu} \rt)
\label {L3}
\end{equation}
in which the Faraday matrix is
\( F_{\mu \nu} = g \, t^a \, F^a_{\mu \nu} \),
the generators \( t^a \)
of \( SU(3) \) are half the Gell-Mann matrices,
and \( F^a_{\mu \nu} = \p_\mu A^a_\nu
- \p_\nu A^a_\mu - 
g f_{a b c} A^b_\mu A^c_\nu \)\@.
The theory described by 
\begin{equation}
L'_3 = m^4 \lt( \frac{1}{\sqrt{1 - L_3/m^4}}
-1 \rt) \quad \mbox{or by} \quad
L''_4 = M^4 \lt( e^{L_3/M^4} -1 \rt)
\label {L'3}
\end{equation}
has Green's functions
that are less singular than those of
the \( L_3 \) theory (\ref{L3})\@.
\par
To simulate such a theory
on a lattice while preserving
gauge invariance,
one may represent the matrix elements
\( A_{\mu \: b c} \)
of the gauge field matrix
\( A_\mu = i \, t^a A_\mu^a \)
in terms of three orthonormal vectors, 
\( e^\dag_b \cdot e_c = \delta_{b c} \),
as inner products of a vector \( e^\dag_b \)
with the derivative \( \p_\mu e_c \)
of another vector~\cite{Ramanan1961, *Ramanan1963, *Cahill11.51}
\begin{equation}
A_{\mu \: b c} =  i \, t^a_{b  c} A_\mu^a
= e^\dag_b \cdot e_{c, \mu}
\equiv e^\dag_b \cdot \p_\mu e_c .
\label {definition of gauge field}
\end{equation}
In this notation,
in which commas denote derivatives,
the elements \( F_{\mu \nu \: a b} \)
of the Faraday matrix are
\beq
F_{\mu \nu \: a b} = \lb D_\mu , D_\nu \rb_{a  b}
= e^\dag_{a , \mu} \cdot e_{b, \nu}
- e^\dag_{a, \mu} \cdot e_c \,\,
e^\dag_c \cdot e_{b, \nu}
- e^\dag_{a, \nu} \cdot e_{b, \mu}
+ e^\dag_{a, \nu} \cdot e_c \,\,
e^\dag_c \cdot e_{b, \mu} .
\label {F^a_{mu nu b} = }
\eeq
This matrix vanishes unless the vectors
have \( n > 3 \) components.
\par
Wilson~\cite{PhysRevD.10.2445}, 
Creutz~\cite{PhysRevLett.43.553,*PhysRevLett.45.313}, 
and others
have demonstrated quark confinement 
on the lattice by replacing the euclidian
action of pure continuum QCD
\begin{equation}
S_{QCD} = \frac{1}{2 g^2} \int 
\tr \lt( F^2_{\mu \nu} \rt) \, d^4x
= \int L_3 \, d^4x 
\label {SQCD}
\end{equation}
by a sum over the 
plaquettes \( \Box \) of a lattice
\begin{equation}
S_W = \sum_\Box S_\Box
\end{equation}
of Wilson's action \( S_\Box \) which 
is the trace of the product
of elements \( U \) of \( SU(3) \) 
on the links that form the plaquette
\begin{equation}
S_\Box = \b \, \lt[ 1 - (1/3)  \re \, \tr \lt(
U_{ij} U_{jk} U_{k \ell} U_{\ell i} \rt) \rt]  .
\label {Sbox}
\end{equation}
\par
Yet there is a big difference between
the continuum action density 
\( \tr ( F^2_{\mu \nu} ) /2g^2 \) 
which can be arbitrarily large 
and Wilson's action \( S_\Box \)
which is bounded by \( \b \)\@. 
This gap is bridged if one uses
the action density \( L'_3  \)
which keeps \( \tr ( F^2_{\mu \nu} ) /2g^2 \) 
bounded like Wilson's \( S_\Box \)\@.
The action density \( L''_3 \) has a similar effect.
Simulations guided by  \( L'_3  \) or \( L''_3 \)
may exhibit ground-state mean values
of the squares of the gauge fields
that are tiny enough to
justify Wilson's compactification
of the gauge fields.
Thus, the ideas of this subsection
may make possible 
a demonstration of quark confinement
without the need to assume that 
compactification is justified.

\subsection{Gravity
\label {Gravity}}

The euclidian action density 
of general relativity is not bounded
below, and so the recipes (\ref{L'}) don't work for it.
Instead, we can use, for instance,
\begin{equation}
L'_E =  \frac{1}{16 \pi \alpha G^2_N} \lt[
 \cosh^2 \th \lt( \frac{ 1 }{(1 - G_N \, R_e)^\alpha} - 1 \rt)
+ \sinh^2 \th  \lt( \frac{ 1 }{(1 + G_N \, R_e)^\alpha} - 1 \rt)
\rt] \sqrt{|g|}
\label {L_E}
\end{equation}
in which \( \alpha > 0 \), \( R_e \) is the euclidian  
Ricci scalar, and \( |g| \)
is the absolute value of the
determinant of the euclidian metric tensor.
We also could use
\begin{equation}
L''_{E} = \frac{1}{16 \pi G^2_N} \lt[
 \cosh^2 \th \lt( e^{G_N R_e} -1 \rt)
 + \sinh^2 \th  \lt( e^{-G_N R_e} -1 \rt) \rt] 
 \sqrt{|g|} .
 \label {L_E'}
\end{equation}
The resulting theories are less singular
than conventional quantum gravity.

\subsection{Fermions
\label{Fermions}}

The energy density of the ground state
of a free Fermi field
is negative and quarticly divergent,
while that of its excited states 
can be arbitrarily high.
So it may make sense
to use a construction similar to 
(\ref{L_E} ) and (\ref{L_E'})\@.
Instead of the usual fermionic action density
\begin{equation}
L_F = \bar \psi (x) ( \gamma_\mu D_\mu + m ) \psi (x) ,
\label {usual fermion action density}
\end{equation}
one could use
\begin{equation}
L'_{F} =  \frac{M^4}{\alpha} \lt[
\cosh^2 \th  \lt( \frac{ 1 }{(1 - L_F/M^4)^\alpha} -1 \rt)
+ \sinh^2 \th \lt( \frac{ 1  }{(1 + L_F/M^4)^\alpha} - 1 \rt) \rt]
\label {L'_F}
\end{equation}
in which \( \alpha > 0 \) or
\begin{equation}
L''_{F} = M^4 \lt[
 \cosh^2 \th \lt( e^{L_F/M^4} -1 \rt)
 + \sinh^2 \th  \lt( e^{-L_F/M^4} -1 \rt) \rt] .
 \label {L''_F}
\end{equation}
These theories are less singular
than the usual theories of fermions.

\section{Summary
\label{Summary}}

Some nonrenormalizable theories
are less singular than every
renormalizable theory.  The space
of such less-singular nonrenormalizable theories is vast.  
Whether any of them 
is realistic or true is unknown.
Two of them, \( L_1 \) (\ref{L_1})
and \( L_s \) (\ref{L_s}), discussed
in section~\ref{Theories with finite Green's functions},
have finite Green's functions and
quadratically divergent ground-state action densities
and describe infinitely massive particles.
Two others,
\( L_f \) (\ref{L_f}) and \( L_{f, s} \) (\ref{L_{f, s}}) of
section~\ref{Theories with finite ground-state action densities},
have finite ground-state action densities
and describe particles of finite mass.
Their bare Green's functions are finite
and vanish in the continuum limit.
Section~\ref{Masses} is about
how I estimated the masses 
of the particles of these four theories.
Section~\ref{Speculations about confinement, gravity, and fermions}
suggests ways of extending the present work
to theories of gauge fields, gravity, and fermions.
Each theory of this paper reduces to its
renormalizable counterpart in the appropriate limit;
for the theories \( L_1 \), \( L_s \), \( L_f \), and \( L_{f, s} \),
that limit is \( M \to \infty \)\@.
The ways of coping with infinities outlined above
apply to theories in any number of space-time dimensions.

\begin{acknowledgments}
I should like to thank 
Jooyoung Lee and the 
Korea Institute for Advanced Study 
for providing computing resources (KIAS Center for Advanced Computation) for this work.
I am grateful to Carl Bender,
J.~Michael Kosterlitz, 
Daniel Topa, David Waxman, 
and Piljin Yi for valuable
suggestions and to David Amdahl, 
Ginette Cahill, 
Luke Caldwell,
John Cherry, 
Alain Comtet, Fred Cooper, 
Michael Creutz,
Michel Dubois-Violette,
Franco Giuliani, Roy Glauber,
Gary Herling, and Tom Hess
for helpful conversations.
\end{acknowledgments}
\bibliography{physics,astro,math}
\end{document}